\documentclass{emulateapj}
\usepackage{graphicx,times}
\usepackage{epstopdf, epsfig}
\usepackage{multirow}
\usepackage{amsmath,amssymb}
\usepackage{rotating}
\usepackage{CJKutf8}

\begin{document}

\title{The Hawaii SCUBA-2 Lensing Cluster Survey: Are Low-luminosity Submillimeter Galaxies Detected in the Rest-frame UV?}

\shortauthors{Hsu et al.}

\author{Li-Yen Hsu \begin{CJK*}{UTF8}{bsmi}(徐立研)\end{CJK*}\altaffilmark{1}, Lennox L. Cowie\altaffilmark{1}, Amy J. Barger\altaffilmark{1,2,3}, and 
Wei-Hao Wang \begin{CJK*}{UTF8}{bsmi}(王為豪)\end{CJK*}\altaffilmark{4}}

\altaffiltext{1}{Institute of Astronomy, University of Hawaii, 2680 Woodlawn Drive, Honolulu, HI 96822, USA}
\altaffiltext{2}{Department of Astronomy, University of Wisconsin-Madison, 475 North Charter Street, Madison, WI 53706, USA}
\altaffiltext{3}{Department of Physics and Astronomy, University of Hawaii, 2505 Correa Road, Honolulu, HI 96822, USA}
\altaffiltext{4}{Academia Sinica Institute of Astronomy and Astrophysics, P.O. Box 23-141, Taipei 10617, Taiwan}

\begin{abstract}

In this third paper of the Hawaii SCUBA-2 Lensing Cluster Survey series, we present Submillimeter Array (SMA) detections of six intrinsically faint 850 $\mu$m sources 
detected in SCUBA-2 images of the lensing cluster fields, A1689, A2390, A370, MACS\,J0717.5+3745, and MACS\,J1423.8+2404. Two of the SCUBA-2 sources split into doublets, 
yielding a total of eight SMA detections. The intrinsic 870 $\mu$m flux densities of these submillimeter galaxies (SMGs) are $\sim$ 1 mJy. Five of the eight SMGs are not detected in optical or near-infrared (NIR) images. The NIR-to-submillimeter flux ratios of these faint SMGs suggest that most of them are extremely dusty and/or at very high redshifts. Combining these 
SMGs and several other samples from the literature, we find a bimodal distribution for the faint sources in the space of submillimeter flux versus NIR-to-submillimeter flux ratio. 
While most of the SMA-detected lensed sources are very obscured, the other SMGs with similar flux densities are mostly bright in the NIR. Future ALMA observations of a 
large sample of SCUBA-2 sources in cluster fields will allow us to decide whether the bimodality 
we observe here really exists.

\end{abstract}

\subjectheadings{cosmology: observations|  galaxies: formation  |  galaxies: starburst  |  gravitational lensing: strong | submillimeter: galaxies }

\section{Introduction}

The measurements of the far-infrared (FIR) Extragalactic Background Light (EBL) demonstrated that about half of the starlight in the optical and ultraviolet (UV) is 
absorbed by dust and re-radiated into the FIR \citep{Puget1996Tentative-detec,Fixsen1998The-Spectrum-of,Dole2006The-cosmic-infr}. It is therefore important 
to study both the unobscured and dust-obscured populations of galaxies across cosmic time for a full picture of star formation in our universe. At high 
redshifts, observations at submillimeter/millimeter wavelengths provide insight on such dusty star formation. However, submillimeter surveys with even 15 m 
telescopes such as the James Clerk Maxwell Telescope (JCMT) become confusion limited \citep{Condon1974Confusion-and-F} at $\lesssim$ 2 mJy at 850 $\mu$m, which 
prevents the detection of fainter submillimeter galaxies (SMGs; see reviews by \citealt{Blain2002Submillimeter-g,Casey2014Dusty-Star-Form}) with infrared 
luminosities $\lesssim10^{12} L_{\odot}$.

Bright SMGs from confusion-limited surveys and the extinction-corrected UV population are essentially disjoint (e.g., \citealt{Barger2014Is-There-a-Maxi,Cowie2017A-Submillimeter}), so 
their contributions to the cosmic star formation history \citep{Madau2014Cosmic-Star-For} must be added. Fainter SMGs, on the other hand, are more common 
objects that contribute the majority of the EBL (e.g., \citealt{Chen2013Resolving-the-C,Hsu2016The-Hawaii-SCUB,Zavala:2017aa}) and therefore most of 
the dusty star formation. However, some of these faint SMGs could also be selected in the UV samples. In order to combine the UV- and FIR-inferred star 
formation history precisely, it is critical to obtain a complete census of faint SMGs that have star formation rates (SFRs) comparable to those of the UV population. Such a 
sample bridges the SFR gap between the two populations and allows us to determine if there is a critical SFR below which UV-selected galaxies alone 
account for all the star formation.

Imaging of massive galaxy cluster fields is a good way to detect intrinsically fainter sources, thanks to gravitational lensing. Lensed sources are magnified at 
all wavelengths, and their images benefit from enhanced spatial resolution. Direct searches for SMGs using interferometry are very inefficient due to the small 
field of views. Thus, deep and wide-field surveys with single-dish submillimeter/millimeter telescopes are the most efficient approach to search for 
SMGs (at least down to their confusion limits).

The SCUBA-2 camera \citep{2013MNRAS.430.2513H} on the JCMT is currently the most powerful instrument for submillimeter surveys. Among all the 
ground-based submillimeter instruments, SCUBA-2 has the best angular resolution ($\sim$ 14$\farcs$5 at 850 $\mu$m and 7$\farcs$5 at 450 $\mu$m), yielding 
more accurate positions, less source blending, and a lower confusion limit. We have been undertaking a SCUBA-2 program, the Hawaii SCUBA-2 Lensing 
Cluster Survey (Hawaii-S2LCS), to image nine massive clusters. These include the northern five clusters in the {\it HST} Frontier Fields 
program \citep{Lotz:2017aa}. We have presented deep number counts at 450 and 850 $\mu$m in \cite{Hsu2016The-Hawaii-SCUB} and a 
radio-detected sample of faint SMGs in \cite{Hsu:2017aa}.

Due to the low spatial resolution of single-dish telescopes, interferometric follow-up is required to identify the multiwavelength counterparts to submillimeter sources. Thus, we have 
also been using the Submillimeter Array (SMA; \citealt{Ho2004The-Submillimet}) to image some of our SCUBA-2 sources. In this third paper of the 
Hawaii-S2LCS series, we present our SMA follow-up observations of six intrinsically faint SCUBA-2 sources discovered in the fields of A1689, A2390, A370, MACS\,J0717.5+3745, 
and MACS\,J1423.8+2404 (hereafter, MACSJ0717 and MACSJ1423). The observations and data reduction are described in Section~2. We present our 
results in Section~3 and discuss their implications in Section~4. Section~5 summarizes this paper. We assume the concordance $\Lambda$CDM cosmology 
with $H_0=70~\rm km~s^{-1}~Mpc^{-1}$, $\Omega_M=0.3$, and $\Omega_\Lambda=0.7$. All magnitudes used are AB magnitudes.

\section{Data}\label{sec:data}

\subsection{SCUBA-2 Observations}

The targets for the SMA observations were selected from our SCUBA-2 lensing cluster surveys, based on the 850 $\mu$m images we had at different times. However, the SCUBA-2 measurements 
we present in this work (Section \ref{sec:SMA_results}) are based on all the 850 $\mu$m data taken with the CV DAISY scan pattern between February 2012 and March 2017. We summarize these 
observations in Table~\ref{tab:table1}. Please refer to \cite{Hsu2016The-Hawaii-SCUB} and \cite{Hsu:2017aa} for details on the data reduction and source extraction procedures. In 
order to correct for the effects of Eddington bias \citep{Eddington1913On-a-formula-fo} and confusion noise \citep{Condon1974Confusion-and-F}, we deboosted the SCUBA-2 flux densities 
based on their signal-to-noise ratios (S/N) using the Monte Carlo simulations described in \cite{Hsu:2017aa}.

\begin{table*}
\caption{Summary of JCMT/SCUBA-2 Observations}
\begin{center}
\begin{tabular}{cccccccc}
\hline \hline	

 Field  & R.A. & Decl. & Redshift  & Weather\footnote{Data were taken in band 1 ($\tau_{\rm 225GHz} < 0.05$), band 2 ($0.05 < \tau_{\rm 225GHz} < 0.08$), or good band 3 ($0.08 < \tau_{\rm 225GHz} < 0.1$) conditions.} & Exposure\footnote{Integration times of the three weather conditions.}   & ${\sigma_{850}}$\footnote{Central 1$\sigma$ sensitivity of the 850 $\mu$m map. These are the statistical/instrumental noise values 
 directly from the reduced rms maps.}    \\
           &       &     &      &        &   (hr)    &   (mJy beam$^{\rm -1}$)    \\
 
\hline

A370                                & 02 39 53.1 & $-$01 34 35.0 & 0.375 &1+2+3 &  28.5+1.5+7.0   &   0.38  \\     

MACSJ0717    & 07 17 34.0 &  \,\,\,\,\,37 44 49.0 & 0.545 &1+2+3 &  32.2+10.5+1.5 & 0.31\\      

A1689                              & 13 11 29.0 & $-$01 20 17.0 & 0.184 & 1+2 & 22.4+1.9  & 0.39 \\     

MACSJ1423     &  14 23 48.3 &  \,\,\,\,\,24 04 47.0 & 0.545 &1+2+3 &  36.5+20.5+1.6 & 0.28  \\      

A2390                              &  21 53 36.8 &  \,\,\,\,\,17 41 44.2 & 0.231 &1+2+3 &   17.4+36.0+9.0 & 0.32\\

\hline \hline  
 
\end{tabular}  
\label{tab:table1}
\end{center}
\end{table*}

\subsection{SMA Observations}

We carried out SMA observations from 2014 to 2016 of six SCUBA-2 850 $\mu$m sources in the five cluster fields. The local oscillator frequency was set at 343 GHz, or 870 $\mu$m. 
The new SWARM (SMA Wideband Astronomical ROACH2 Machine) correlator and dual receiver mode became available during the course of our program, which 
greatly improved the continuum sensitivity. We summarize these observations in Table~\ref{tab:table2}.

We used the SMA data reduction package MIR to calibrate our data. The visibilities were first weighted in inverse proportion to the square of the system temperatures. The continuum data 
were generated by averaging all the spectral channels after performing passband phase calibration. We used the gain calibrators to correct for the variations of phase/amplitude in time, and 
then we performed the flux calibration to set the absolute flux level. For the track executed in dual receiver mode (20161017 in Table~\ref{tab:table2}), we ran all the calibrations for the two receivers 
separately.

The visibilities from all the available tracks for each source were combined and imaged using the interferometry data reduction package MIRIAD \citep{Sault:1995aa}. We made 
the dirty maps and the synthesized dirty beam images in a 0$\farcs$2 (compact configuration) or 0$\farcs$1 (extended configuration) grid using the routine {\sc invert} with natural 
weighting on the baselines. We also performed multi-frequency synthesis, which gives better coverage in the frequency-dependent uv coordinate. The {\sc clean} routine was used 
to deconvolve the dirty map. We {\sc clean}ed the images around detected sources to approximately 1.5$\,\sigma$ to remove the effects of sidelobes. The resulting source 
fluxes are not sensitive to the depth to which we chose to clean. Primary beam correction was applied to the images by dividing the {\sc clean}ed fluxes by the off-axis gain. In 
Table~\ref{tab:table3}, we summarize the synthesized beams and central sensitivities of the final images.

\begin{table*}
\begin{center}
\caption{Summary of SMA Observations}

\begin{tabular}{ccccccccc}
\hline \hline	

 ID  & Field  & Configuration & Track Dates & Receiver(s) & Bandwidth\footnote{Total bandwidth combing upper and lower sidebands. When the SWARM correlator 
 operated, an additional 2 GHz was available for each sideband.} & Passband    & Gain             &       Flux        \\
      &            &                     &                        &                  &                  & Calibrator     & Calibrator(s)  &  Calibrator(s)\\

\hline

SMA-1 & A370              & extended & 20151001  & 345  & 8 GHz & 3c454.3 &  0224+069, 0309+104 & Neptune \\
            &                       &                 & 20161017 & 345, 400\footnote{Dual receiver mode. The two receivers cover the same spectral range.} & 12 GHz & 3c454.3 & 0224+069, 0309+104 & Uranus\\ 
\hline
SMA-2 & MACSJ0717  & compact  & 20160102  &  345 & 8 GHz & 3c84     & 0818+423, 0927+390 & Neptune \\
\hline
SMA-3 & A1689            &                 & 20160104   &  345 & 12 GHz & 3c279 & 1337-129, 3c279 & Callisto \\ 
\hline
SMA-4 & MACSJ1423  & compact   & 20160301 &  345 & 8 GHz & 3c273   & 1504+104 & Ganymede\\ 
            &                       &                  & 20160308 &  345 & 8 GHz & 3c273    & 1357+193, 1415+133 & Ganymede\\
\hline
SMA-5 & A2390           & compact    & 20140625  & 345 & 8 GHz  & 3c279 & 2148+069, 3c454.3 & Titan, Neptune \\
            &                      &                   & 20140626 &  345 & 8 GHz  & 3c279  & 2148+069, 3c454.3 & Neptune \\
            &                      &                   & 20141024 &  345 & 8 GHz  & 3c454.3  & 2148+069, 3c454.3 & Neptune \\ 
            &                      &                   & 20141029 &  345 & 8 GHz  &  3c454.3  & 2148+069, 3c454.3 & Neptune \\ 
            &                      &                   & 20141030 &  345 & 8 GHz  &  3c454.3  & 2148+069, 3c454.3 & Neptune \\ 
\hline
SMA-6 & A2390          & compact     & 20140629 & 345  & 8 GHz  &  3c279  & 2148+069, 3c454.3 & Titan, Neptune \\
            &                     &                    & 20140630 & 345 & 8 GHz  &   3c279  & 2148+069, 3c454.3 & Titan, Neptune \\
            &                     &                    & 20140702 & 345 & 8 GHz  &   3c454.3  & 2148+069, 3c454.3 & Titan, Neptune \\
            &                     &                    & 20141024 & 345 & 8 GHz  &   3c454.3  & 2148+069, 3c454.3 & Neptune \\ 
            &                     &                    & 20141029 & 345 & 8 GHz  &   3c454.3  & 2148+069, 3c454.3 & Neptune \\

\hline \hline  
 
\end{tabular}  

\label{tab:table2}
\end{center}
\end{table*}

\begin{table}
\caption{Synthesized Beam Sizes and Position Angles as well as Central Sensitivities of the SMA Images}
\begin{center}
\begin{tabular}{cccc}
\hline \hline	

 ID  &    Beam FWHM & Beam P.A. & $\sigma$ \\
      &    ($'' ~\times~ ''$) &   (deg)        &   (mJy beam$^{-1}$) \\

\hline
SMA-1 &   0.86 $\times$ 0.60 & \,\,\,\,\,85.0 &  0.44\\
SMA-2 &   2.20 $\times$ 1.88 & $-$64.6 & 0.38 \\
SMA-3 &   2.03 $\times$ 1.92 & $-$0.9 & 0.50 \\ 
SMA-4 &   2.15 $\times$ 1.78 & $-$86.3 & 0.48 \\ 
SMA-5 &   2.28 $\times$ 1.60 & $-$73.9 & 0.55 \\
SMA-6 &   2.21 $\times$ 1.53 & $-$83.0 & 0.43 \\

\hline \hline  
 
\end{tabular}  
\label{tab:table3}
\end{center}
\end{table}

\subsection{HST and Spitzer Images}

For the Frontier Field clusters, we retrieved the Advanced Camera for Surveys (ACS) and Wide-Field Camera 3 (WFC3) images from the {\it HST} Frontier Field 
archive\footnote{https://archive.stsci.edu/pub/hlsp/frontier/}. The images for MACSJ1423 are taken from the Cluster Lensing And Supernova survey with Hubble 
(CLASH; \citealt{Postman2012The-Cluster-Len}) archive\footnote{https://archive.stsci.edu/prepds/clash/}. For A1689 and A2390, we used SWARP \citep{Bertin:2002aa} 
to combine individual archival images\footnote{PI (PIDs): Blakeslee (11710), Ellis (10504), Ford (9289, 11802), Rigby (11678), Siana (12201, 12931)} for each passband. 
We also retrieved the {\it Spitzer} Frontier Fields data\footnote{http://irsa.ipac.caltech.edu/data/SPITZER/Frontier/}, as well as the ``Super Mosaics'' and their source 
catalogs for the other three clusters from the {\it Spitzer} archive.

\subsection{$K_{s}$-band Images}\label{sec:K-band}

We carried out $K_{s}$-band observations of A1689 and A2390 (PI: Hsu; PID: 15AH83) with WIRCam on the Canada--France--Hawaii Telescope (CFHT) 
in 2015. Along with the archival data of A2390 (PI: Umetsu; PID: 07AT98), the total integration times are 2800 and 3515 seconds for A1689 and A2390, respectively. 
We reduced and combined these images using Imaging and Mosaicking PipeLinE (SIMPLE; \citealt{Wang:2010aa}), an IDL-based package for 
galactic/extragalactic imaging with CFHT/WIRCam.

SIMPLE performs flat fielding, background subtraction, distortion correction, absolute astrometry, photometric calibration, wide-field mosaicking, cosmic ray removal, and image weighting. Absolute 
astrometry was obtained by comparing the image with the source catalogs of the Sloan Digital Sky Survey (SDSS). The photometry was calibrated with bright stars in the source catalogs 
of the Two Micron All Sky Survey (2MASS). We reduced the data chip by chip before mosaicking, and the pixel scale of the images is 0$\farcs$3. The images reach 3$\,\sigma$ limiting 
magnitudes of 23.8 and 24.0 in a $1''$-radius aperture for A1689 and A2390, respectively.

We also retrieved the $K_s$-band images of the Frontier Fields A370 (VLT/HAWK-I), MACSJ0717 (Keck/MOSFIRE), and MACSJ1423 (CFHT/WIRCam) from 
\cite{Brammer2016Ultra-deep-K-S-} and the CLASH archive, respectively.

\subsection{VLA Images}\label{sec:VLA}

We make use of the 3 GHz image of MACSJ0717 taken with the Karl G. Jansky Very Large Array (VLA) from \cite{Hsu:2017aa}, which has a synthesized beam of 
$\sim$ 0$\farcs$4 and a sensitivity of $\sim$ 1 $\mu$Jy. For A370 and A2390, we obtained the VLA 1.4 GHz images from \cite{Wold2012Very-Large-Arra}. The 
A370 (A2390) image has a synthesized beam of $\sim$ 1$\farcs$7 (1$\farcs$4) and a noise level of $\sim$ 5.7 (5.6) $\mu$Jy near the cluster center.

\section{Results}\label{sec:SMA_results}

\subsection{SMA Detections and Multiwavelength Counterparts}\label{sec:counterpart}

We detected eight sources above a 4$\,\sigma$ level in the six SMA images, where SMA-2 and SMA-3 both split into doublets. The positions and flux densities of the original SCUBA-2 sources 
and these SMA detections are summarized in Table~\ref{tab:table4}. In Figure~1, we show the postage stamp images centered at the original SCUBA-2 positions. Although there is more than a 
factor of 2 discrepancy between the SCUBA-2 and SMA flux densities for SMA-1 and SMA-6, the difference can be caused by multiple faint sources that are below our detection limit. This is clear for SMA-6, where we can see some emission with S/N $> 3$ coming from an optically detected galaxy (possibly a pair of interacting galaxies). 

Three (SMA-1, SMA-2-1, and SMA-5) of the eight SMA sources are detected in the optical or NIR. Note that for SMA-4, there is one elliptical galaxy that has a 0$\farcs$87 offset from the SMA position. Given the positional uncertainties of the SMA and {\it HST} detections, the offset is measured above a 5$\,\sigma$ level. In addition, based on the photometric catalog from CLASH, this elliptical galaxy is a cluster 
member at $z \sim 0.5$. Thus, it is unlikely to be the counterpart to the SMG. There is also a small offset between the peaks of the submillimeter and optical emission for SMA-5. However, the offset becomes smaller in the $K_{s}$-band image. We suggest that the optical/NIR source is the correct counterpart to the SMG, and the offset is due to different regions of unobscured and dust-obscured star formation, which is very typical for SMGs.

As we will discuss in Section~\ref{sec:discuss}, we use $K_s$-band photometry to quantify the NIR emission, given that $K_s$-band images are available for all of our sources. In addition, they 
have better spatial resolution than the {\it Spitzer} images. We used SExtractor \citep{Bertin1996SExtractor:-Sof} to measure {\sc auto} magnitudes of the SMA sources detected at $K_s$ band. The 
deblending parameters {\sc deblend\_nthresh} and {\sc deblend\_mincont} were set to be 32 and 0.005, respectively. For the sources that are not detected in the $K_s$-band 
images, we measured their 3$\,\sigma$ limiting magnitudes in a $1''$-radius aperture.

\begin{figure*}
\begin{center}
\includegraphics[width=3.6cm]{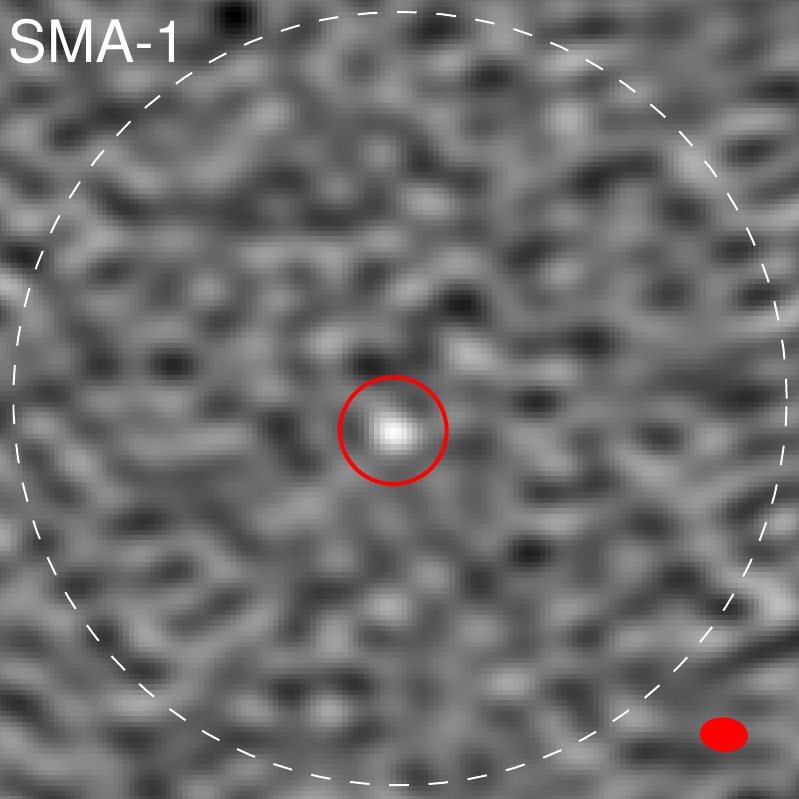}
\includegraphics[width=3.6cm]{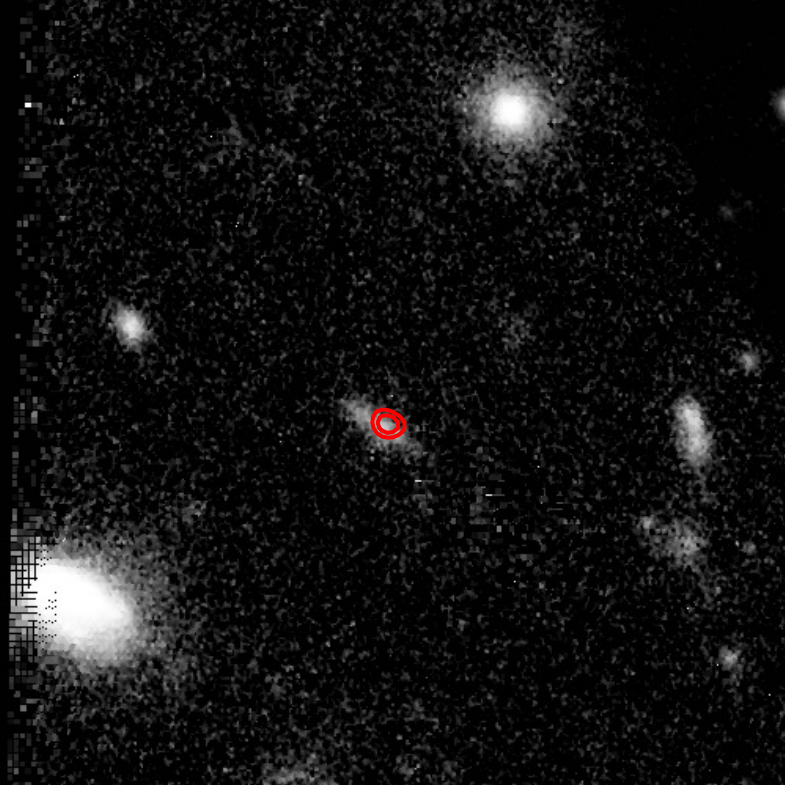} 
\includegraphics[width=3.6cm]{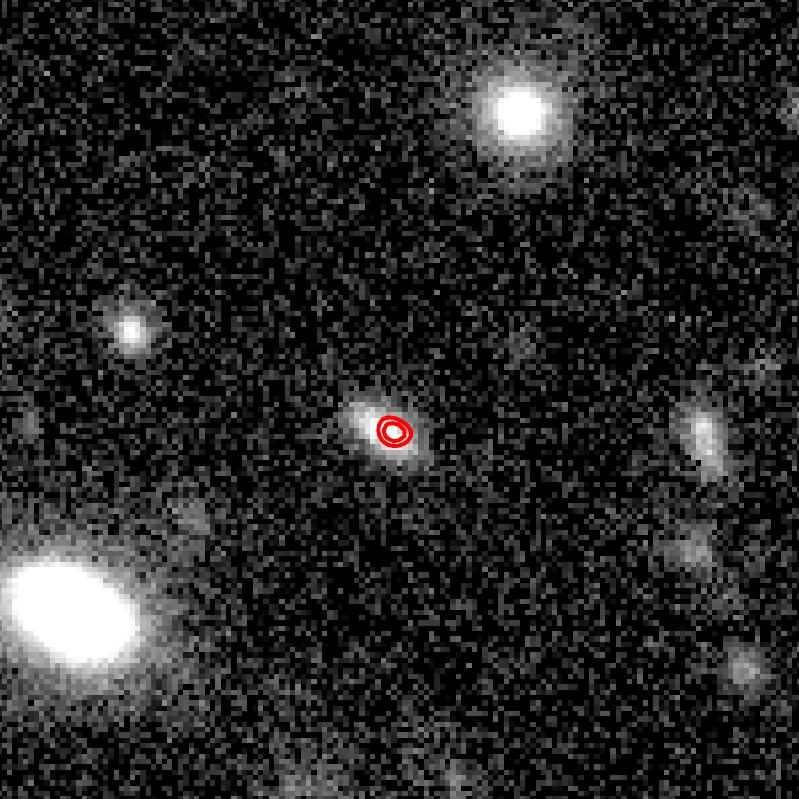}
\includegraphics[width=3.6cm]{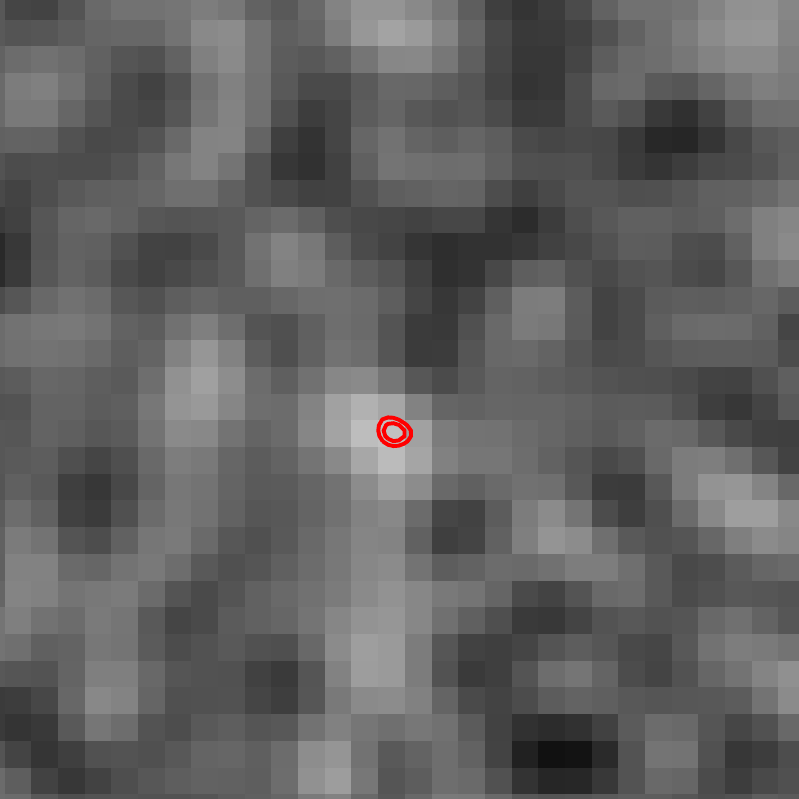} \\ [0.3mm] 

\includegraphics[width=3.6cm]{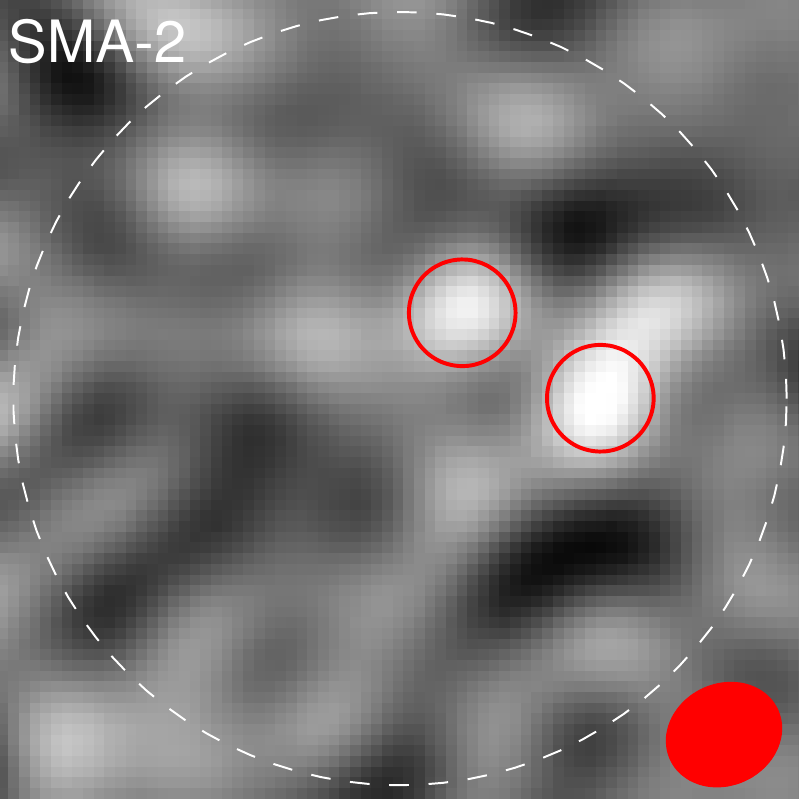}
\includegraphics[width=3.6cm]{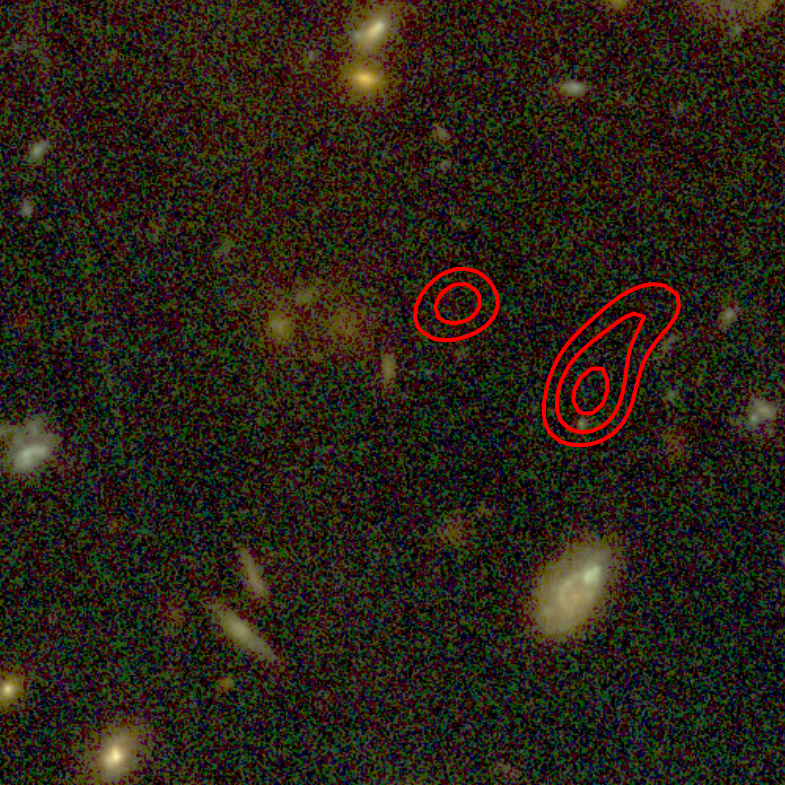} 
\includegraphics[width=3.6cm]{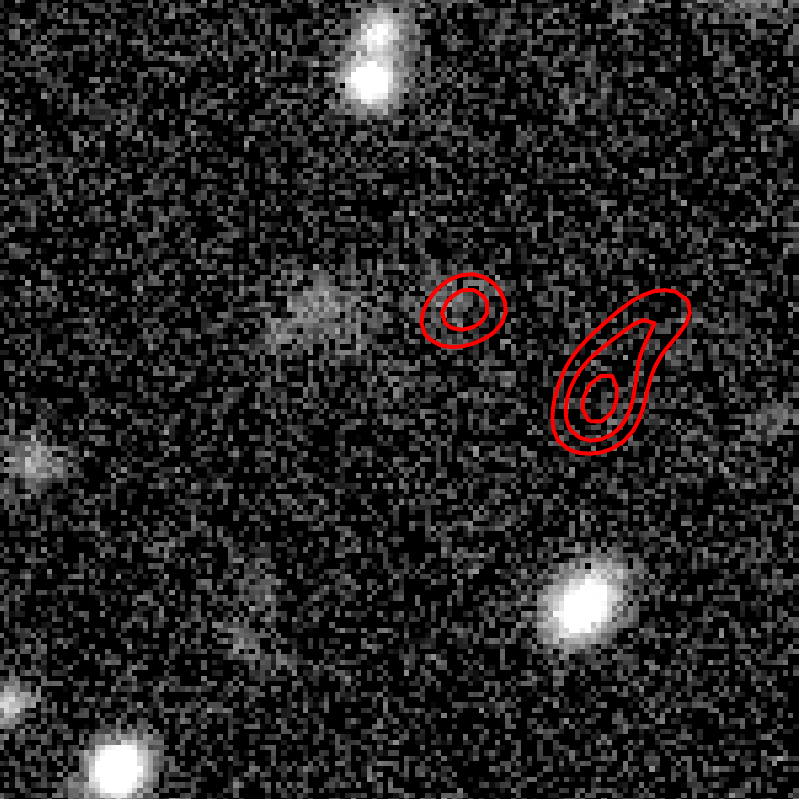}
\includegraphics[width=3.6cm]{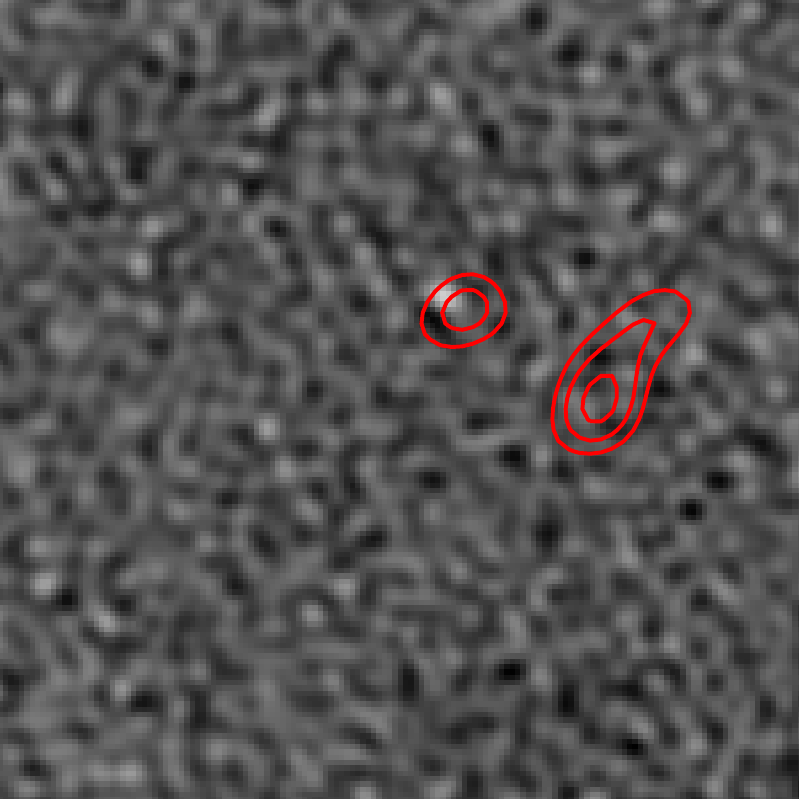} \\ [0.3mm]

\hspace{-3.66cm}\includegraphics[width=3.6cm]{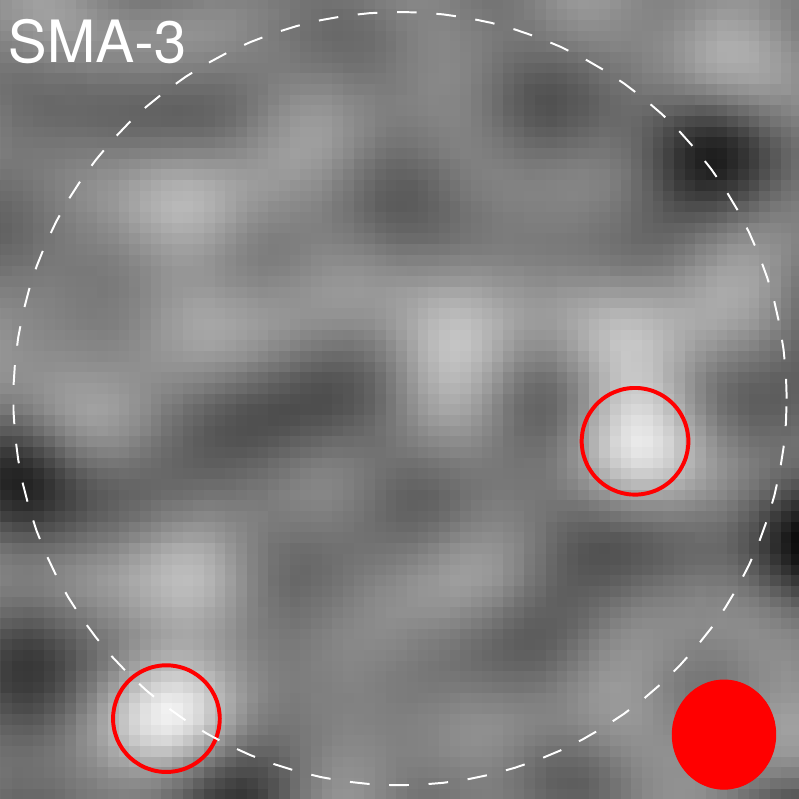}
\includegraphics[width=3.6cm]{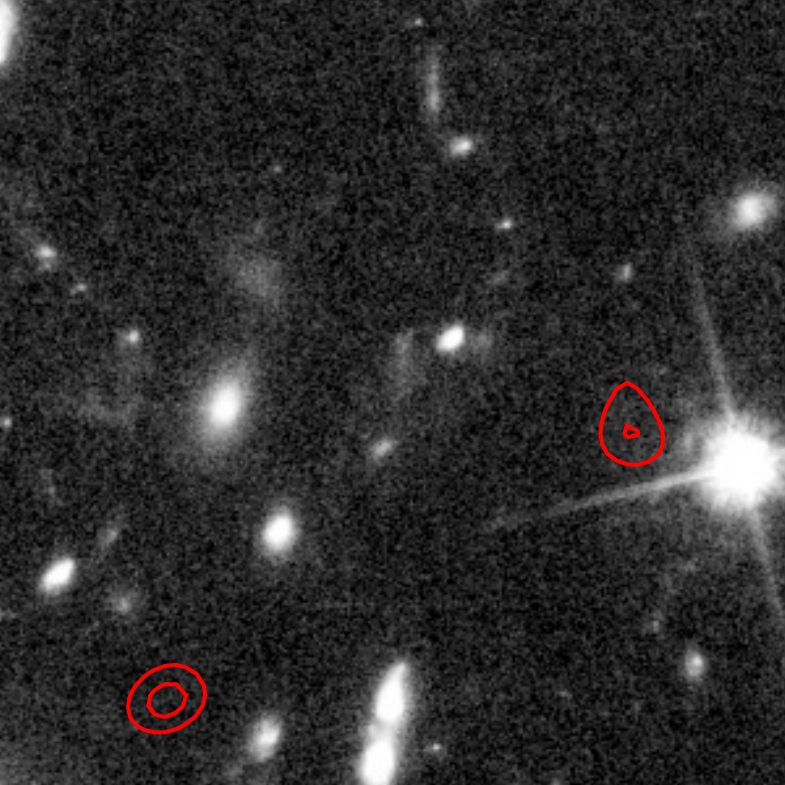} 
\includegraphics[width=3.6cm]{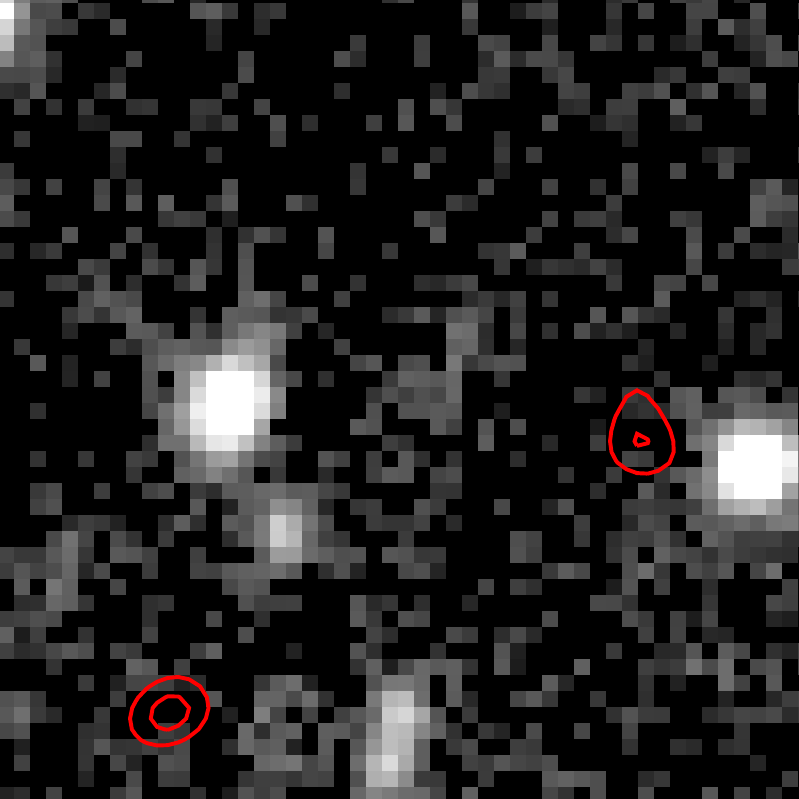} \\ [0.3mm]

\hspace{-3.66cm}\includegraphics[width=3.6cm]{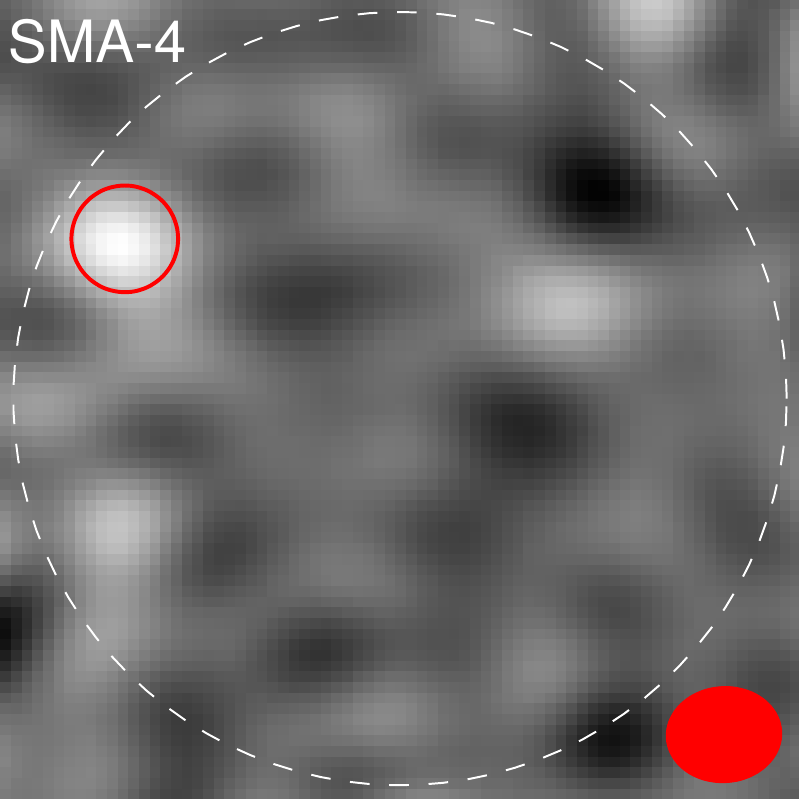}
\includegraphics[width=3.6cm]{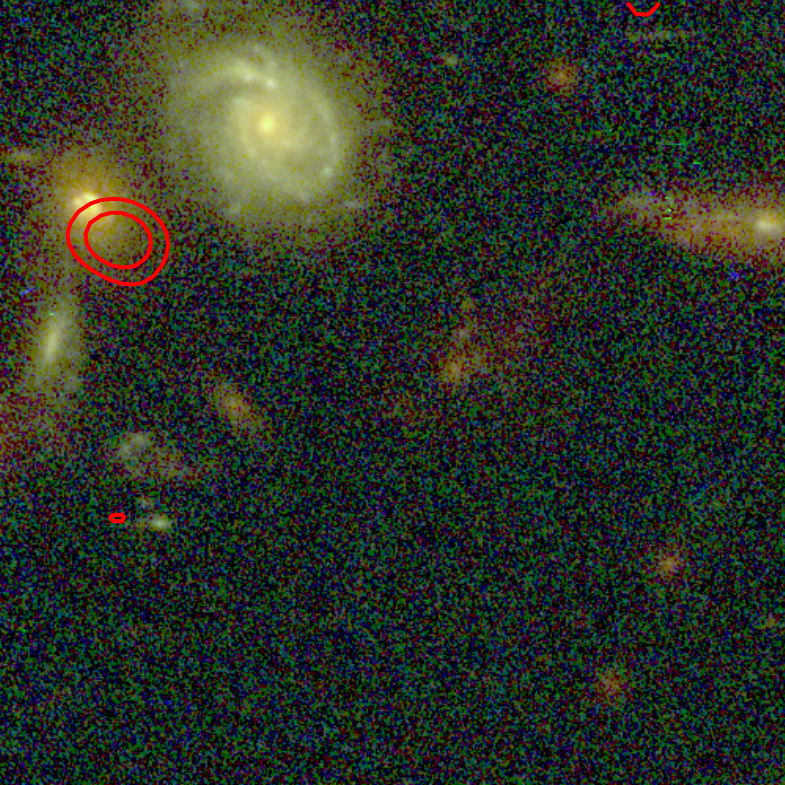} 
\includegraphics[width=3.6cm]{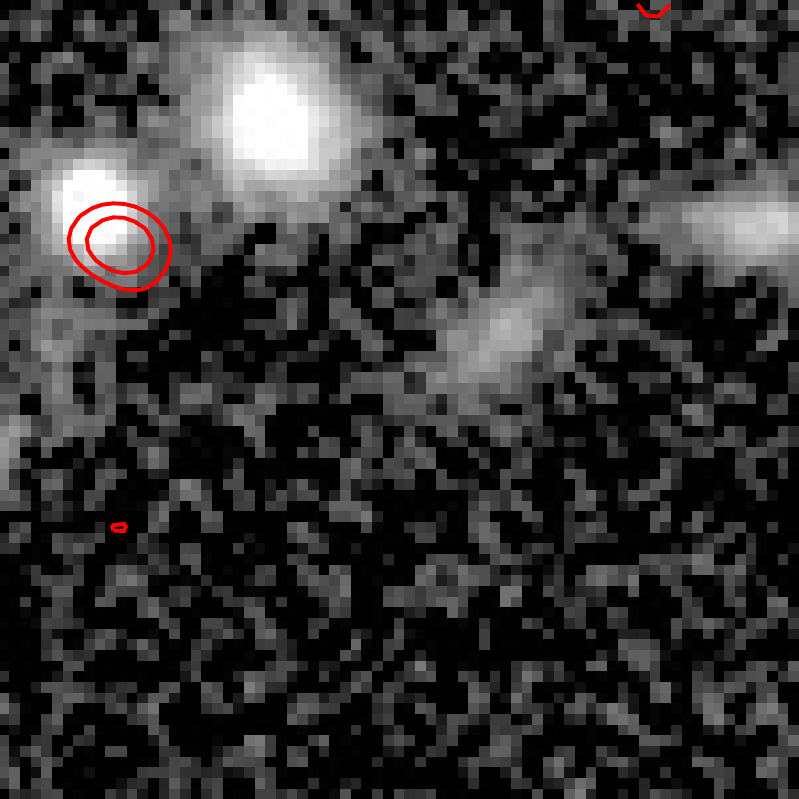}\\ [0.3mm]

\includegraphics[width=3.6cm]{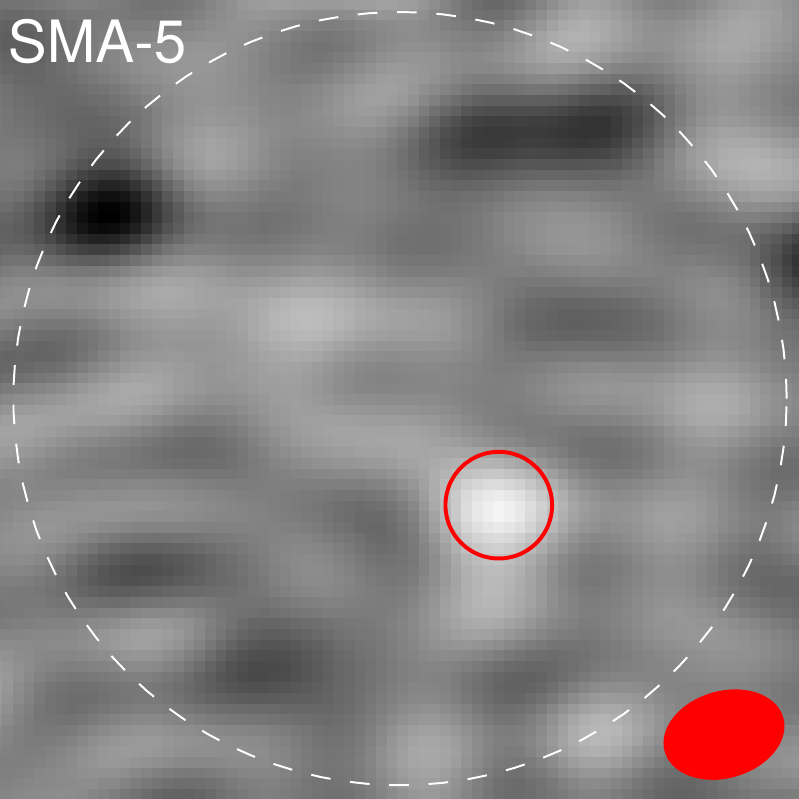}
\includegraphics[width=3.6cm]{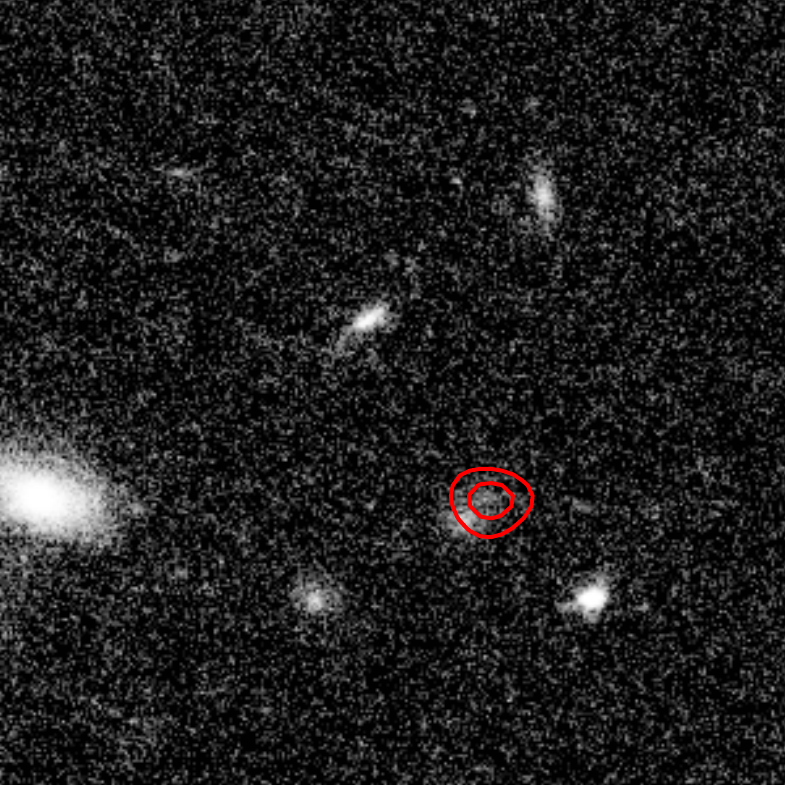} 
\includegraphics[width=3.6cm]{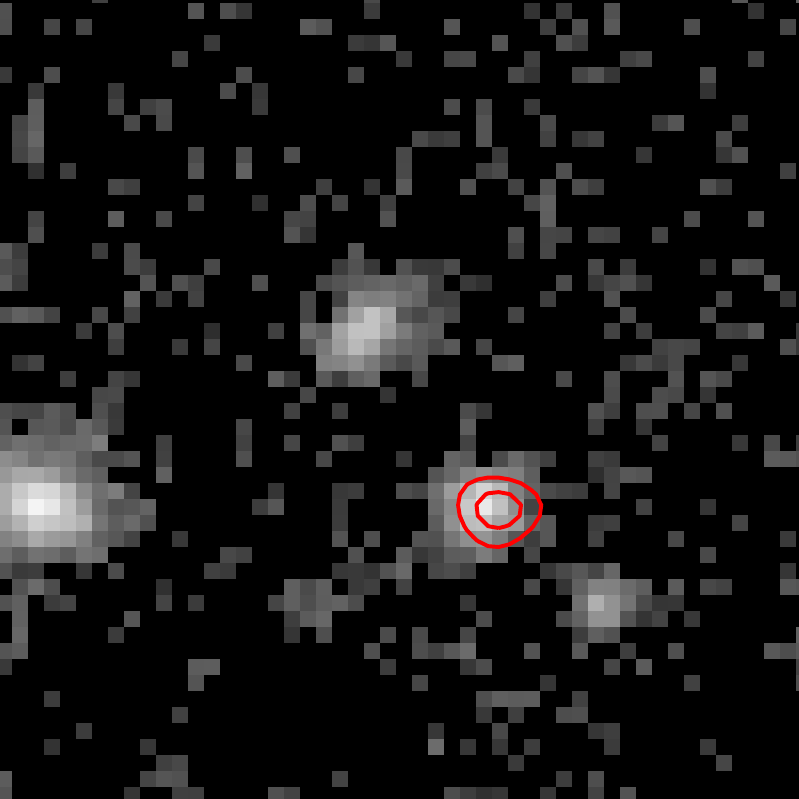}
\includegraphics[width=3.6cm]{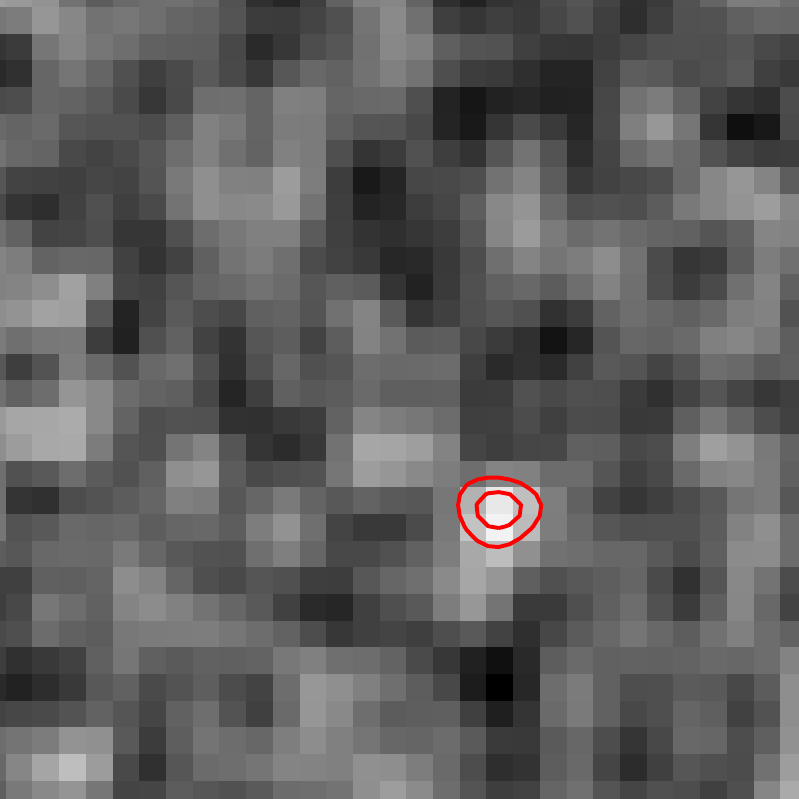} \\ [0.3mm]

\includegraphics[width=3.6cm]{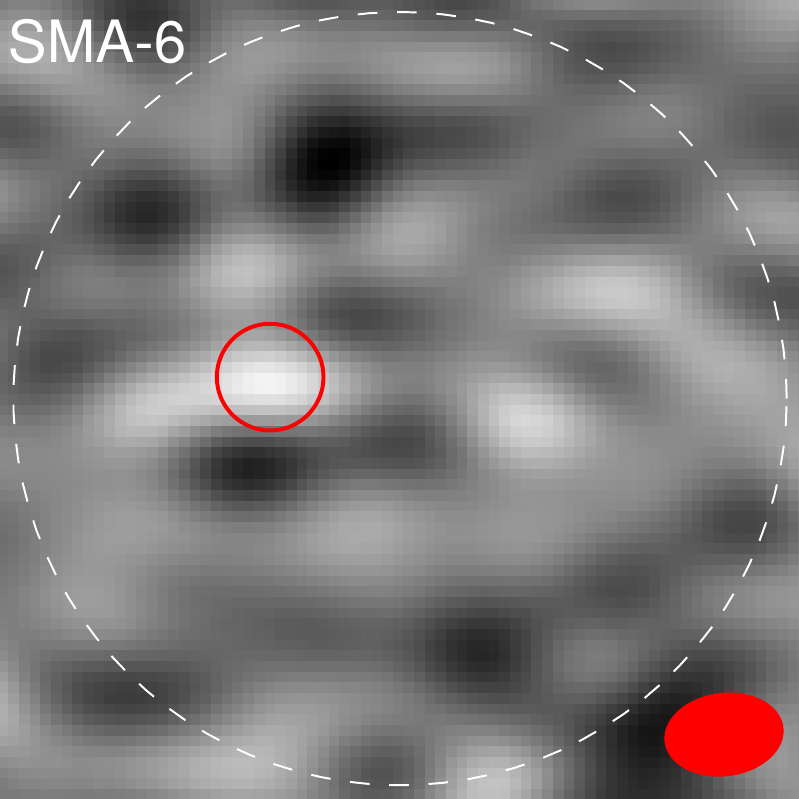}
\includegraphics[width=3.6cm]{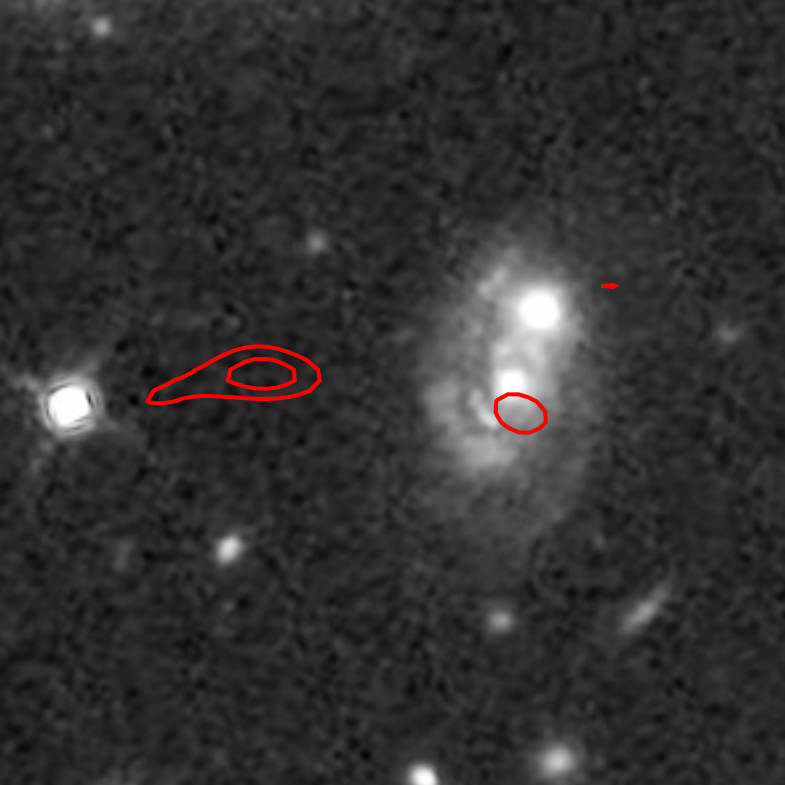} 
\includegraphics[width=3.6cm]{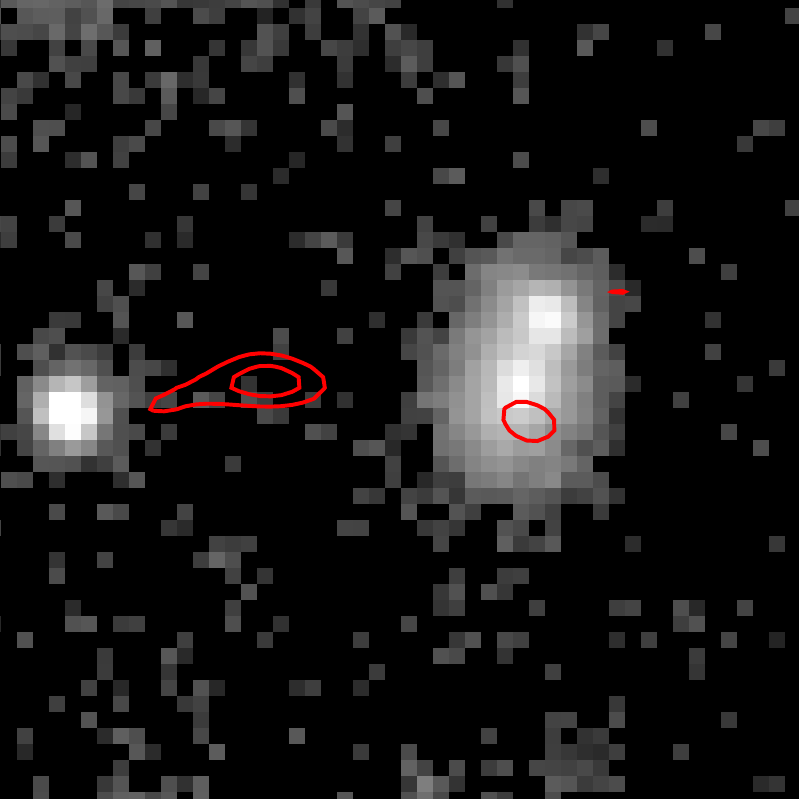}
\includegraphics[width=3.6cm]{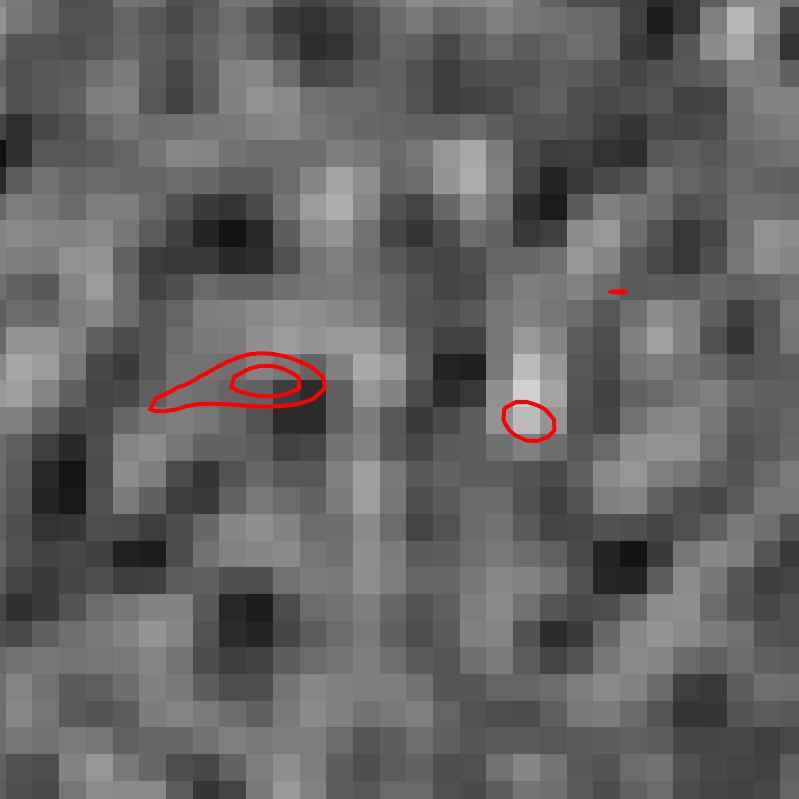} 

\caption[Multiwavelength images for the SMA-detected SMGs]{Postage stamp images for the SMA detections centered at the SCUBA-2 850 $\mu$m positions. From left to right are SMA 870 $\mu$m, {\it HST} (from top to bottom: F160W, F435W-F606W-F814W false color, F814W, F435W-F606W-F814W false color, F850LP, and F125W), $K_s$-band (SMA-1: VLT/HAWK-I, SMA-2: 
Keck/MOSFIRE; SMA-3, SMA-4, SMA-5, and SMA-6: CFHT/WIRCam), and VLA (SMA-1, SMA-5, and SMA-6: 1.4 GHz; SMA-2: 3 GHz) 
images. The image size is 15$'' \times$ 15$''$. In the SMA images, the large dashed circles with a diameter of 14$\farcs$5 represent the JCMT beam (FWHM); we 
use 1$''$-radius red circles to denote the SMA detections, and the ellipses at the bottom-right corners represent the synthesized beams. The red contours in the 
other images are (3, 4, 5)$\times \sigma$ isophotes of the SMA sources.}

\label{fig:figure1}
\end{center}
\end{figure*}

\subsection{Redshift Estimates}\label{sec:redshift}

We measured the photometric redshifts of SMA-1 and SMA-5 using the BPZ code (Bayesian photometric redshift estimation; \citealt{Benitez2000Bayesian-Photom}) and the \cite{Bruzual2003Stellar-populat} models 
with the \cite{Chabrier2003Galactic-Stella} initial mass function. SMA-1 is covered by {\it HST} observations at F435W, F606W, F814W, F110W, F140W, and F160W. We ran SExtractor in dual-image 
mode using F160W as the detection band to obtain {\sc auto} magnitudes. The deblending parameters {\sc deblend\_nthresh} and {\sc deblend\_mincont} were again chosen to be 32 and 0.005, respectively.  For 
SMA-5, F850LP is the only available {\it HST} passband. We took the IRAC $1.9''$-radius aperture photometry from the {\it Spitzer} source catalog of A2390 and ran SExtractor in single-image mode to 
measure the {\sc auto} magnitudes at F850LP and $K_{s}$. Before running BPZ, we corrected all the magnitudes for Galactic dust extinction from \cite{Schlafly2011Measuring-Redde}. We obtained 
$z = 2.39 \pm 0.17$ and $2.00 \pm 0.15$ for SMA-1 and SMA-5, respectively. 

For SMA-2-1, SMA-2-2 and  SMA-6, we used the submillimeter-to-radio flux ratios to compute their ``millimetric redshifts'', following the method in \cite{Barger2000Mapping-the-Evo}. The 
relation between the redshift and the submillimeter-to-radio flux ratio is
\begin{equation}
\label{eq:milli}
\begin{split}
z & = (S_{343{\rm GHz}}/S_{1.4{\rm GHz}})^{0.26} -1 \\ 
   & = 0.85(S_{343{\rm GHz}}/S_{3{\rm GHz}})^{0.26} -1
\end{split}
\end{equation}
With $S_{3{\rm GHz}} = 5.13 \pm 1.51$ $\mu$Jy \citep{Hsu:2017aa}, we obtained $z = 2.9 \pm 0.4$ for SMA-2-1. Because SMA-2-2 and SMA-6 are not detected in the radio images, we 
used their 3$\,\sigma$ limits, 3.05 $\mu$Jy (3 GHz) and 22.6 $\mu$Jy (1.4 GHz), to compute the lower redshift limits. The results are $z > 3.7$ and $z > 2.2$ for SMA-2-2 and SMA-6, respectively. 
Note that SMA-1 and SMA-5 are also detected at 1.4 GHz with $S_{1.4{\rm GHz}} = 26.9 \pm 6.2$ $\mu$Jy and $35.8 \pm 6.9$ $\mu$Jy, respectively. Their millimetric redshifts are 
2.1 $\pm$ 0.3 and 2.0 $\pm$ 0.2, respectively, in agreement with the photometric redshifts.

For the three SMA sources in A1689 and MACSJ1423, there are no deep radio images available. Because these sources are not detected in {\it HST}, $K_s$-band, or {\it Spitzer} images, we expect them to be at high redshifts. We assume a conservative lower limit at $z = 1.0$ to estimate their lensing magnifications, which we will describe in the next section.

\begin{table*}
\begin{center}
\caption{Positions and Flux Densities of the SCUBA-2 Sources and their SMA Detections}

\begin{tabular}{ccccccc}
\hline \hline	

        &              & SCUBA-2  &          &         &  SMA  &                   \\

 ID  &  R.A.  &  Decl.  & $S_{850}$ & R.A.  &  Decl.  & $S_{870}$ \\
       &           &            &    (mJy)      &          &            &     (mJy)       \\
\hline

SMA-1 & 02 39 57.57 & $-$01 34 53.0 & 4.71 $\pm$ 0.73 & 02 39 57.58 & $-$01 34 53.6 &  2.14 $\pm$ 0.44\\

SMA-2    & 07 17 38.22 & \,\,\,\,\,37 46 15.0  & 2.97 $\pm$ 0.63 &  ... & ... & ... \\
SMA-2-1 & ...  & ...  & ...  &  07 17 38.12  &  \,\,\,\,\,37 46 16.6  &1.78 $\pm$ 0.39\\
SMA-2-2 & ...  & ...  & ...  &  07 17 37.90  &  \,\,\,\,\,37 46 15.0 &2.15 $\pm$ 0.40\\

SMA-3    & 13 11 23.93 & $-$01 20 46.4 & 4.26 $\pm$ 0.77 &  ... & ... & ... \\
SMA-3-1 & ... & ... & ... &  13 11 23.64  &  $-$01 20 47.2   & 2.14 $\pm$ 0.53\\
SMA-3-2 & ... & ... & ... &  13 11 24.22  &  $-$01 20 52.4   & 2.49 $\pm$ 0.57\\

SMA-4 &  14 23 48.14 &  \,\,\,\,\,24 04 11.1   &   2.70 $\pm$ 0.51 &  14 23 48.52   &  \,\,\,\,\,24 04 14.1     & 2.64 $\pm$ 0.54\\
SMA-5 &  21 53 34.63 &  \,\,\,\,\,17 40 31.2 &  3.39 $\pm$ 0.64 & 21 53 34.50 &  \,\,\,\,\,17 40 29.2   & 2.49 $\pm$ 0.56\\ 
SMA-6 &  21 53 38.69 &  \,\,\,\,\,17 42 17.2 &  4.01 $\pm$ 0.66 & 21 53 38.86 &  \,\,\,\,\,17 42 17.6   &1.97 $\pm$ 0.43\\

\hline \hline  
 
\end{tabular}  
\label{tab:table4}
\end{center}
\end{table*}

\begin{table*}
\begin{center}
\caption{Lens Models Used for Each Cluster Field} 

\begin{tabular}{ccc}
\hline \hline	
 Field  &  Models \\ 
\hline 

{A370}    & Bradac-v1~~CATS-v1~~Merten-v1~~Sharon-v2~~Williams-v2\\
            &        Zitrin-LTM-v1~~Zitrin-LTM-Gauss-v1~~Zitrin-NFW-v1 \\  

{MACSJ0717} & Bradac-v1~~CATS-v4.1~~Diego-v4.1~~GLAFIC-v3~~Keeton-v4~~Merten-v1\\
                     & Sharon-v4~~Williams-v4~~Zitrin-LTM-v1~~Zitrin-LTM-Gauss-v1 \\                 
        
A1689   &  \cite{Limousin2007Combining-Stron} \\   
MACSJ1423    &   Zitrin-LTM-Gauss-v2~~Zitrin-NFW-v2 \\           
A2390   &  \cite{Richard2010LoCuSS:-first-r} \\

\hline \hline
\end{tabular}  
\label{tab:table5}
\end{center}
\end{table*}

\subsection{Lens Models}\label{sec:lens}

In order to compute the magnifications and intrinsic flux densities of our faint SMGs, the lens models of the clusters and source redshifts are required. A set of lens models 
are available for the {\it HST} Frontier Fields from ten teams, including Bradac \citep{Bradac2005Strong-and-weak,Bradac2009Focusing-Cosmic,Hoag2016The-Grism-Lens-}, Caminha \citep{Caminha2017}, CATS \citep{Jullo2009Multiscale-clus,Jauzac2012A-weak-lensing-,Jauzac2014Hubble-Frontier,Jauzac2015MNRAS.452.1437,Jauzac2015Hubble-Frontier,2014MNRAS.444..268R}, Diego \citep{Diego2005Non-parametric-,Diego2005-A1689,Diego2007Combined-recons,Diego2015}, GLAFIC \citep{Oguri2010The-Mass-Distri,Kawamata2016Precise-Strong-}, Keeton \citep{Keeton:2010aa,Ammons:2014aa,McCully:2014aa}, Merten \citep{Merten2009Combining-weak-,Merten2011Creation-of-cos}, Sharon \citep{Jullo2007A-Bayesian-appr,Johnson2014Lens-Models-and}, Williams \citep{Liesenborgs2006A-genetic-algor,Mohammed2014,Grillo2015CLASH-VLT:-Insi,Sebesta2016}, and Zitrin \citep{Zitrin2009New-multiply-le,Zitrin2013CLASH:-The-Enha}.

For A1689, A2390, and MACSJ1423, we used the models from \cite{Limousin2007Combining-Stron}, \cite{Richard2010LoCuSS:-first-r}, and the CLASH archive, respectively. Both the 
Frontier Fields and CLASH archives provide a set of images to account for the full range (i.e., the uncertainty) of each model, and we used the newest model from each team. On 
the other hand, only the best-fit models\footnote{https://projets.lam.fr/projects/lenstool/wiki} are available for A1689 and A2390 by running the LENSTOOL software \citep{Kneib1996Hubble-Space-Te}. 
In Table~~\ref{tab:table5}, we tabulate the models we used for each cluster field.

Following \cite{Coe2015Frontier-Fields}, we estimated the median and 68.3\% range of the magnification values from Monte Carlo simulations. For each source, we propagated the positional and redshift 
uncertainties, as well as the full range (except for A1689 and A2390) of all the available lens models. The positional uncertainties were measured using the MIRIAD {\sc imfit} routine, and 
the typical values are 0$\farcs$1$\sim$0$\farcs$3 in both right ascension and declination. To propagate the redshift uncertainties of SMA-2-1, SMA-2-2, SMA-3-1, SMA-3-2, SMA-4, and SMA-6, we 
used a uniform distribution between their lower limits and an upper limit at $z = 6$. The resulting magnification error for SMA-4 is very large, because the critical lines at $z=1-6$ are close to 
this source. We therefore decided to only use $z > 1.0$ to compute the lower limit of its magnification. In Table~~\ref{tab:table6}, we summarize the redshifts, lensing magnifications, de-lensed 
submillimeter flux densities, and observed $K_s$-band magnitudes of the SMA sources.

\begin{table}
\begin{center}
\caption{Redshifts, Lensing Magnifications, De-lensed Submillimeter Flux Densities, and Observed $K_s$-band Magnitudes of 
the SMA Sources}

\begin{tabular}{ccccc}
\hline \hline	

 ID  &  $z$  &  $\mu$  & $S_{870,{\rm int}}$ & $m_{K_s}$ \\
       &          &              &    (mJy)                   &    (mag)    \\
\hline
SMA-1    & $2.39 \pm 0.17$  & $1.88^{+0.35}_{-0.36}$ & $1.14 \pm 0.32 $           & 22.5 \\
SMA-2-1 & $2.9 \pm 0.4$      & $1.86^{+0.26}_{-0.49}$ & 0.96$^{+0.33}_{-0.25}$ & 24.4 $\pm$ 0.2 \\ 
SMA-2-2 &  $>3.7$               & $2.09^{+0.34}_{-0.59}$ & 1.03$^{+0.35}_{-0.25}$ & $> 25.4$ \\ 
SMA-3-1 &  ...               & $2.81^{+0.10}_{-0.20}$ & 0.76$^{+0.20}_{-0.19}$ & $> 23.8$ \\
SMA-3-2 &  ...               & $3.31^{+0.15}_{-0.35}$ & 0.75$^{+0.19}_{-0.18}$ & $> 23.8$ \\ 
SMA-4    &  ...               & $> 1.96$                        & $< 1.35$                        & $> 23.3$ \\
SMA-5    &  $2.00 \pm 0.15$ & $ 2.09 \pm 0.02 $          & $1.19 \pm 0.27$            & 21.4 $\pm$ 0.2 \\ 
SMA-6    &   $>2.2$              & $3.96^{+0.12}_{-0.20}$ & $0.50 \pm 0.11$            & $> 24.0$ \\
\hline \hline  
 
\end{tabular}  
\label{tab:table6}
\end{center}
\end{table}

\begin{figure*}
\begin{center}
\includegraphics[width=12.0cm]{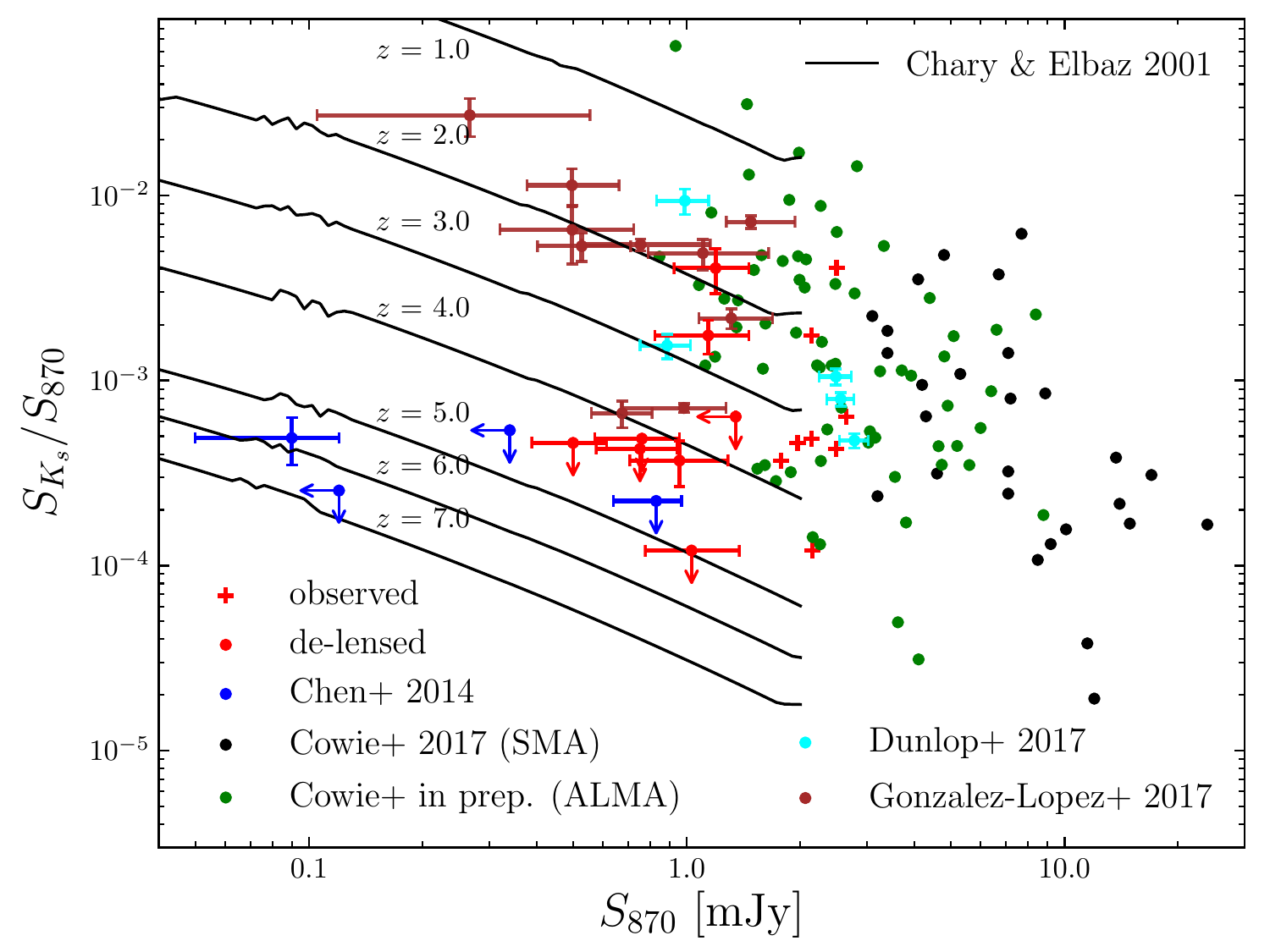}
\caption[$K_s$-to-870 $\mu$m flux ratios versus 870 $\mu$m flux densities for (sub)millimeter-identified SMGs]{$K_s$-to-870 $\mu$m flux ratios versus 870 $\mu$m 
flux densities of our SMA-detected SMGs (red crosses: observed; red circles: de-lensed) and other samples. Blue circles represent SMA-detected lensed SMGs in A1689 and 
A2390 from \cite{Chen2014SMA-Observation}. Cyan circles are ALMA-detected 1.3 mm sources in HUDF from \cite{Dunlop2017}, and we scaled their 
flux densities to 870 $\mu$m values assuming an Arp 220 SED \citep{Silva1998Modeling-the-Ef}. ALMA-detected lensed SMGs from \cite{Gonzalez-Lopez2017The-ALMA-Fronti} 
are shown in brown, where we again used an Arp 220 SED to scale their 1.1 mm flux densities to 870 $\mu$m values. The flux densities of \cite{Chen2014SMA-Observation} 
and \cite{Gonzalez-Lopez2017The-ALMA-Fronti} are corrected for lensing magnifications. Black circles are SMA-detected bright SMGs in CDF-N from \cite{Cowie2017A-Submillimeter}. 
ALMA-detected SMGs in CDF-S (Cowie et. al., in preparation) are shown in green. The predictions based on the SED templates of \cite{Chary2001Interpreting-th} at various 
redshifts are plotted in black curves. There is a trend of increasing $K_s$-to-870 $\mu$m flux ratio with decreasing flux density for most of the sources. However, the majority of 
SMA-detected lensed SMGs do not follow the same trend. }

\label{fig:figure2}
\end{center}
\end{figure*}

\begin{figure*}
\begin{center}
\includegraphics[width=12.0cm]{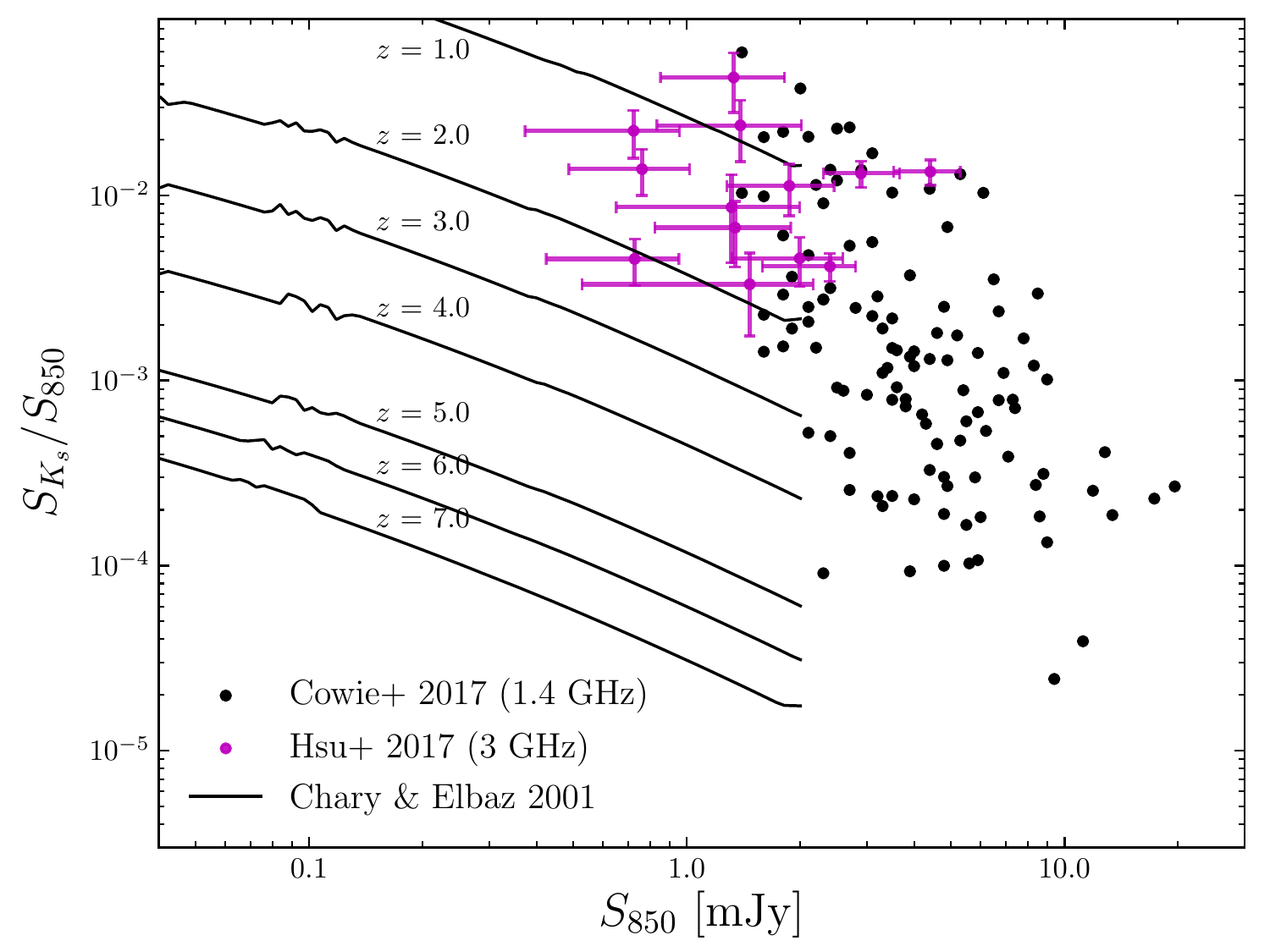}
\caption[$K_s$-to-850 $\mu$m flux ratios versus 850 $\mu$m flux densities for radio-identified SMGs]{$K_s$-to-850 $\mu$m flux 
ratios versus SCUBA-2 850 $\mu$m flux densities of the bright SMGs identified with VLA 1.4 GHz in CDF-N (black) from \cite{Cowie2017A-Submillimeter} and the lensed 
SMGs identified with VLA 3 GHz (purple) from \cite{Hsu:2017aa}. Note that one of the 14 sources (0717--2) in \cite{Hsu:2017aa} is removed because it corresponds to 
SMA-2-1 in this work. The predictions based on the SED templates of \cite{Chary2001Interpreting-th} at various redshifts are plotted in black curves.}

\label{fig:figure3}
\end{center}
\end{figure*}

\begin{figure}
\begin{center}
\includegraphics[width=6cm]{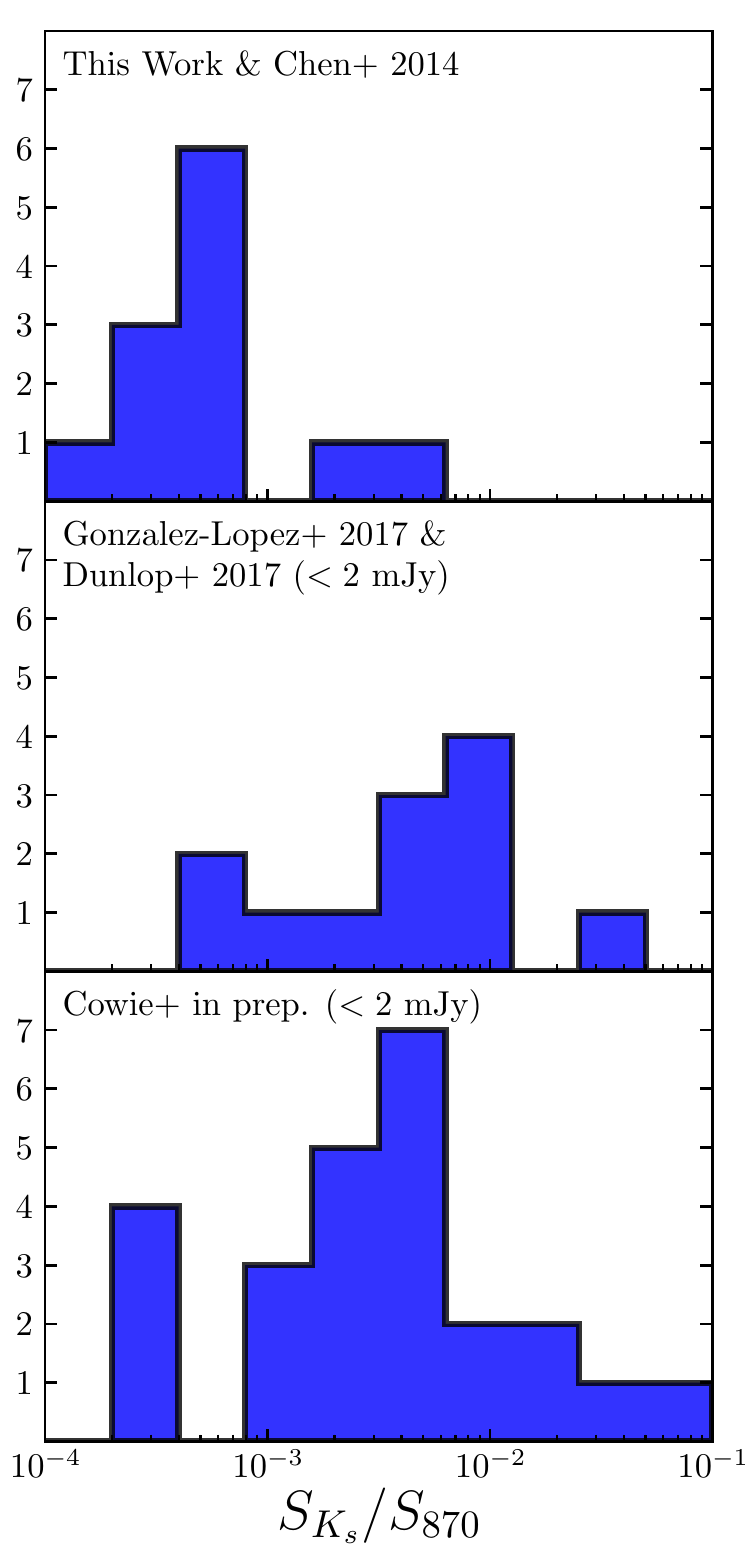}

\caption[Histograms of the $K_s$-to-870 $\mu$m flux ratios]{Histograms of the $K_s$-to-870 $\mu$m flux ratios for various samples of faint SMGs. Top: 
SMA-detected sources from this work and \cite{Chen2014SMA-Observation}. Middle: \cite{Gonzalez-Lopez2017The-ALMA-Fronti} and the sources 
that are fainter than 2 mJy from \cite{Dunlop2017}, with the flux densities scaled to 870 $\mu$m values. Bottom: the sources that are fainter than 2 mJy 
from Cowie et. al. (in preparation). The median (de-lensed) 870 $\mu$m flux densities for these samples are 0.80, 0.82, and 1.54 mJy, respectively. A K-S test 
suggests that the top and middle samples are not drawn from the same distribution.}

\label{fig:figure4}
\end{center}
\end{figure}

\section{Discussion}\label{sec:discuss}

Five of our eight SMA sources are not detected in either the optical or NIR images. This agrees with \cite{Chen2014SMA-Observation} and suggests that many faint SMGs are still missed by 
optical surveys and would not be included in the UV star formation history. However, studies of low-redshift starburst galaxies (e.g., \citealt{Chary2001Interpreting-th,Le-Floch:2005aa,Reddy2010Dust-Obscuratio}) 
have shown that fainter sources are generally less dusty. In addition, some recent work suggests that fainter SMGs are on average at lower redshifts (e.g., \citealt{Heavens2004The-star-format,Bundy2006The-Mass-Assemb,Franceschini2006Cosmic-evolutio,Dye2008The-SCUBA-HAlf-,Mobasher2009Relation-Betwee,Magliocchetti2011The-PEP-survey:,Hsu2016The-Hawaii-SCUB,Cowie2017A-Submillimeter}). Based 
on these results, the NIR-to-submillimeter flux ratios of SMGs are expected to increase with decreasing luminosity/flux.

Recently, we have obtained different samples of SCUBA/SCUBA-2 sources followed up by the SMA \citep{Chen2014SMA-Observation,Cowie2017A-Submillimeter}, ALMA (Cowie et. al., in preparation), or 
VLA \citep{Cowie2017A-Submillimeter,Hsu:2017aa}. Since $K_s$-band imaging is available for these samples and this work, we can combine them and inspect the change of $K_s$-to-submillimeter flux 
ratio over a wide flux range. In Figure~\ref{fig:figure2} (\ref{fig:figure3}), we show $K_s$-to-870 (850) $\mu$m flux ratio as a function of 870 (850) $\mu$m flux density for the submillimeter (radio) 
identified samples. The 1.1 mm lensed sources in the Frontier Fields from \cite{Gonzalez-Lopez2017The-ALMA-Fronti} and the 1.3 mm sources in the Hubble Ultra Deep Field (HUDF; \citealt{Beckwith:2006aa}) 
from \cite{Dunlop2017}\footnote{The final sample of \cite{Dunlop2017} consists of sixteen 3.5$\,\sigma$-detected sources that have NIR counterparts in the {\it HST} image. This is a clean but 
biased sample, since any real source without a NIR counterpart is rejected. We therefore only include their five 6$\,\sigma$-detected sources, which comprise a clean and unbiased sample 
without the need of detections in the NIR.} are also included in Figure~\ref{fig:figure2}. We scaled their flux densities to 870 $\mu$m values using an Arp 220 spectral energy distribution (SED; \citealt{Silva1998Modeling-the-Ef}) redshifted to their 
redshifts \citep{Laporte:2017aa,Dunlop2017}. For example, the conversions are $S_{870\mu{\rm m}}/S_{1.1{\rm mm}} = $ 1.92 and $S_{870\mu{\rm m}}/S_{1.3{\rm mm}} = $ 3.10 for a source at $z =2$. For the $K_s$-band photometry\footnote{Two of the twelve 1.1 mm sources from \cite{Gonzalez-Lopez2017The-ALMA-Fronti}, A2744-ID02 and A2744-ID03, are seriously blended in the $K_s$-band image, so we did not 
measure their photometry. These two sources are therefore not included in Figures~\ref{fig:figure2} and \ref{fig:figure4}. \cite{Laporte:2017aa} simply used a 
0$\farcs$4-radius aperture with the IRAF NOAO daophot package and applied aperture corrections to measure the magnitudes for 11 of these sources. However, here 
we performed our own measurements with SExtractor instead of taking their results.}, we ran SExtractor on the images 
from \cite{Brammer2016Ultra-deep-K-S-} and \cite{Fontana:2014aa} for the 1.1 and 1.3 mm sources, respectively.

We can see in Figures~\ref{fig:figure2} and \ref{fig:figure3} that most of the SMGs show a trend of increasing $K_s$-to-submillimeter flux ratios as we go from brighter to 
fainter sources. The 3 GHz-identified lensed SMGs from \cite{Hsu:2017aa} and the faintest sources in the CDF-S ALMA sample (Cowie et. al., in preparation) occupy roughly the same space 
of the diagrams. However, the majority of SMA-detected lensed SMGs from \cite{Chen2014SMA-Observation} and this work do not seem to be drawn from the same 
population to which the other samples belong. In Figure~\ref{fig:figure4}, we compare the distributions of the $K_s$-to-870 $\mu$m flux ratios 
for the SMA-detected lensed SMGs from this work and \cite{Chen2014SMA-Observation}, the ALMA-detected lensed 1.1 mm sources \citep{Gonzalez-Lopez2017The-ALMA-Fronti} 
and 1.3 mm sources \citep{Dunlop2017} that are fainter than 2 mJy at 870 $\mu$m, and the ALMA-detected blank-field SMGs (Cowie et. al., in preparation) that are fainter than 2 mJy. 
The median (de-lensed) 870 $\mu$m flux densities for these three samples are 0.80, 0.82, and 1.54 mJy, respectively. A K-S test for the first and the second samples results in a 
$p$-value $ < 0.001$ and therefore suggests that they are not drawn from the same distribution. Note that SMA-1 and SMA-5, the two sources that have the highest $K_s$-to-870 $\mu$m 
flux ratios among our SMA sample (see Figures~\ref{fig:figure2} and~\ref{fig:figure4}), are both detected at 1.4 GHz above a 4$\,\sigma$ level.

We can see a bimodal color distribution in the left side of Figure~\ref{fig:figure2}. This suggests that besides optically bright and/or low-redshift sources, there is a 
population of faint SMGs that are extremely dusty and/or at very high redshifts. However, based on the K-S test, there might be a selection bias in our SMA 
samples. These sources were chosen for SMA observations because they were candidates to be highly magnified SMGs, especially for the ones of 
\cite{Chen2014SMA-Observation}. ALMA imaging of our SCUBA-2 sources in the cluster centers will be the best approach to obtain a large and even 
sample of faint SMGs. Given the efficiency of ALMA observations, other pre-selections based on magnifications or observed flux densities are 
not required. As a consequence, we will be able to decide whether the bimodality we observe here really exists.

\section{Summary}

We carried out SMA observations of six intrinsically faint 850 $\mu$m sources detected by SCUBA-2 in lensing cluster fields, A1689, A2390, A370, MACS\,J0717.5+3745, and 
MACS\,J1423.8+2404, yielding a total of eight SMA detections. Two of the SCUBA-2 sources split into doublets. Based on the lens models from the literature, the intrinsic 
870 $\mu$m flux densities of these SMGs are $\sim$ 1 mJy. Five of the sources have no optical or NIR counterparts. The NIR-to-submillimeter flux ratios of 
these faint SMGs suggest that most of them are extremely dusty and/or at very high redshifts. Combining this work and several other samples of SMGs identified with 
ALMA or SMA, we found a bimodal distribution for the faint sources in the space of submillimeter flux versus NIR-to-submillimeter flux ratio. However, there might be 
a selection bias in the SMA-detected lensed sources (this work and \citealt{Chen2014SMA-Observation}). Future ALMA observations of a large sample of SCUBA-2 
sources in cluster fields will allow us to decide whether the bimodality we observe here really exists.

 \acknowledgments

We gratefully acknowledge support from NSF grants AST-0709356 (L.-Y. H. and L. L. C.) and AST-1313150 (A. J. B.), as well as the John Simon Guggenheim Memorial Foundation and Trustees of the William F. Vilas Estate (A. J. B.). James Clerk Maxwell Telescope is operated by the East Asian Observatory on behalf of The National Astronomical Observatory of Japan, Academia Sinica Institute of Astronomy and Astrophysics, the Korea Astronomy and Space Science Institute, the National Astronomical Observatories of China and the Chinese Academy of Sciences (Grant No.\,XDB09000000), with additional funding support from the Science and Technology Facilities Council of the United Kingdom and participating universities in the United Kingdom and Canada. The James Clerk Maxwell Telescope has historically been operated by the Joint Astronomy Centre on behalf of the Science and Technology Facilities Council of the United Kingdom, the National Research Council of Canada and the Netherlands Organisation for Scientific Research. Additional funds for the construction of SCUBA-2 were provided by the Canada Foundation for Innovation. The National Radio Astronomy Observatory is a facility of the National Science Foundation 
operated under cooperative agreement by Associated Universities, Inc. We acknowledge the cultural significance that the summit of Mauna Kea has to the indigenous Hawaiian community.
 
\bibliographystyle{apj}
\bibliography{ms}

\end{document}